\documentclass[twocolumn]{emulateapj}

\usepackage{subfigure}

\usepackage{bm}

\usepackage{amsmath, amssymb}
\usepackage{times}

\usepackage[usenames]{xcolor}
\definecolor{mpl_red}{HTML}{D62728}

\usepackage[backref, breaklinks, plainpages=false, colorlinks=true, anchorcolor=blue!50!black, citecolor=blue!50!black, linkcolor=blue!50!black, urlcolor=mpl_red, bookmarks=false]{hyperref}
\usepackage[strict]{changepage}

\usepackage{soul}
\usepackage{upgreek}

\begin{document}

\renewcommand*{\backref}[1]{[#1]}

\newcommand{\Majid}{\href{https:/orcid.org/0000-0002-4694-4221}{\textcolor{blue!50!black}{Walid~A.~Majid}}}
\newcommand{\Pearlman}{\href{https:/orcid.org/0000-0002-8912-0732}{\textcolor{blue!50!black}{Aaron~B.~Pearlman}}}
\newcommand{\Prince}{\href{https:/orcid.org/0000-0002-8850-3627}{\textcolor{blue!50!black}{Thomas~A.~Prince}}}
\newcommand{\Naudet}{\href{https://orcid.org/0000-0001-6898-0533}{\textcolor{blue!50!black}{Charles~J.~Naudet}}}
\newcommand{\Wharton}{\href{https:/orcid.org/0000-0002-7416-5209}{\textcolor{blue!50!black}{Robert~S.~Wharton}}}
\newcommand{\Bansal}{\href{https:/orcid.org/0000-0002-7418-7862}{\textcolor{blue!50!black}{Karishma~Bansal}}}

\newcommand{\Connor}{\href{https://orcid.org/0000-0002-7587-6352}{\textcolor{blue!50!black}{Liam~Connor}}}

\newcommand{\Bhardwaj}{\href{https:/orcid.org/0000-0002-3615-3514}{\textcolor{blue!50!black}{Mohit~Bhardwaj}}}

\newcommand{\Tendulkar}{\href{https:/orcid.org/0000-0003-2548-2926}{\textcolor{blue!50!black}{Shriharsh~P.~Tendulkar}}}



\newcommand{\JPL}{Jet Propulsion Laboratory, California Institute of Technology, Pasadena, CA 91109, USA; \textcolor{blue}{walid.majid@jpl.nasa.gov}}
\newcommand{\CaltechPhysics}{Division of Physics, Mathematics, and Astronomy, California Institute of Technology, Pasadena, CA 91125, USA}
\newcommand{\McGill}{Department of Physics, McGill University, 3600 rue University, Montréal, QC H3A 2T8, Canada}
\newcommand{\MSI}{McGill Space Institute, McGill University, 3550 rue University, Montréal, QC H3A 2A7, Canada}

\newcommand{\MSIFellow}{$^{\text{6}}$~McGill Space Institute~(MSI) Fellow.}
\newcommand{\FRQNTFellow}{$^{\text{7}}$~FRQNT Postdoctoral Fellow.}
\newcommand{\NPPFellow}{$^{\text{8}}$~NASA Postdoctoral Program Fellow.}

\newcommand{\Dunlap}{Dunlap Institute for Astronomy \& Astrophysics, University of Toronto, 50 St.~George Street, Toronto, ON M5S 3H4, Canada}
\newcommand{\Toronto}{David A.~Dunlap Department of Astronomy \& Astrophysics, University of Toronto, 50 St.~George Street, Toronto, ON M5S 3H4, Canada}
\newcommand{\MITKavli}{MIT Kavli Institute for Astrophysics and Space Research, Massachusetts Institute of Technology, 77 Massachusetts Ave, Cambridge, MA 02139, USA}
\newcommand{\MITPhysics}{Department of Physics, Massachusetts Institute of Technology, 77 Massachusetts Ave, Cambridge, MA 02139, USA}
\newcommand{\TATA}{Department of Astronomy and Astrophysics, Tata Institute of Fundamental Research, Mumbai, 400005, India}
\newcommand{\UBC}{Department of Physics \& Astronomy, University of British Columbia, 6224 Agricultural Road, Vancouver, BC V6T 1Z1, Canada}
\newcommand{\API}{Anton Pannekoek Institute for Astronomy, University of Amsterdam, Science Park 904, 1098 XH, Amsterdam, The Netherlands}



\journalinfo{{\sc Accepted for Publication in The Astrophysical Journal Letters}}

\shorttitle{A Bright Fast Radio Burst with Sub-100-Nanosecond Structure}
\shortauthors{MAJID ET AL.}

\title{A Bright Fast Radio Burst from FRB~20200120E with Sub-100-Nanosecond Structure}

\author{\Majid\altaffilmark{1,2}, \Pearlman\altaffilmark{2,3,4,6,7}, \Prince\altaffilmark{1,2}, \Wharton\altaffilmark{1,8}, \Naudet\altaffilmark{1}, \Bansal\altaffilmark{1},
\Connor\altaffilmark{2},
\Bhardwaj\altaffilmark{3,4}, 
and~\Tendulkar\altaffilmark{5}
}

\address{
$^{\text{1}}$~\JPL \\
$^{\text{2}}$~\CaltechPhysics \\
$^{\text{3}}$~\McGill \\
$^{\text{4}}$~\MSI \\
$^{\text{5}}$~\TATA
}

\thanks{\MSIFellow}
\thanks{\FRQNTFellow}
\thanks{\NPPFellow}


\begin{abstract}
\label{Section:Abstract}

We have detected a bright radio burst from FRB~20200120E  with the NASA Deep Space Network~(DSN) 70\,m dish (DSS-63) at radio frequencies between 2.2–-2.3\,GHz. 
This repeating fast radio burst~(FRB) is reported to be associated with a globular cluster in the M81 galactic system. 
With high time resolution recording, low scattering, and large intrinsic brightness of the burst, we find a burst duration of $\sim$30\,$\mu$s, comprising of several narrow components with typical separations of 2--3\,$\mu$s.
The narrowest component has a width of $\lesssim 100$~ns, which corresponds to a light travel time size as small as 30\,m. 
The peak flux density of the narrowest burst component is 270\,Jy. 
We estimate the total spectral luminosity of the narrowest component of the burst to be 4\,$\times$\,10$^{\text{30}}$\,erg\,s$^{\text{-1}}$\,Hz$^{\text{-1}}$, which is a factor of $\sim$500 above the luminosities of the so-called ``nanoshots'' associated with giant pulses from the Crab pulsar.
This spectral luminosity is also higher than that of the radio bursts detected from the Galactic magnetar SGR~1935+2154 during its outburst in April 2020, but it falls on the low-end of the currently measured luminosity distribution of extragalatic FRBs, 
further indicating the presence of a continuum of FRB luminosities. 
The temporal separation of the individual components has similarities to the quasi-periodic behavior seen in the microstructure of some pulsars.  The known empirical relation between the microstructure quasi-periodicity timescale and the  rotation period of pulsars possibly suggests a possible pulsar as the source of this FRB, with a rotation period of a few milliseconds.

\end{abstract}



\section{Introduction}
\label{Section:Introduction}

\setcounter{footnote}{8}

Fast radio bursts (FRBs) are energetic, short duration radio transients that are highly dispersed, with  dispersion measures~(DMs) that are well in excess of the expected Galactic contribution along their line of sights (see, e.g.~\citealt{Petroff+2019, Cordes+2019} for recent reviews). The localization of a subset of FRBs to host galaxies at redshifts of 0.034--0.66 has confirmed the extragalactic nature of FRBs~\citep{Chatterjee+2017, Bannister+2019, Ravi+2019, Prochaska+2019, Marcote+2020}. The current census of FRBs includes over 100 sources~\citep{Petroff+2016}\footnote{See \href{https://www.herta-experiment.org/frbstats/catalogue}{https://www.herta-experiment.org/frbstats/catalogue.}},
24 of which have been found to produce repeat bursts (e.g.,~see~\citealt{Spitler+2016, CHIME+2019a, CHIME+2019, Fonseca+2020}). Despite the increasing number of FRB detections, their origin remains an open question with numerous progenitor and emission models that have been proposed to explain the nature of these mysterious bursts (see, e.g.~\citealt{Platts+2019}\footnote{See \href{http://frbtheorycat.org}{http://frbtheorycat.org.}} for a catalog of theories and proposed models). Follow-up studies of repeaters with broadband instruments at higher time and frequency resolution as well as detailed polarimetric studies of repeating bursts are critical for constraining the progenitor models and for understanding of the FRB phenomenon (see, e.g., \citealt{Nimmo+2021, Nimmo+2021b}).

FRB~20200120E is a repeating FRB, recently discovered by the Canadian Hydrogen Intensity Mapping Experiment Fast Radio Burst (CHIME/FRB) instrument \citep{Bhardwaj+2021}.
FRB~20200120E has the lowest reported dispersion measure (DM)  to date among FRBs (87.82\,pc\,cm$^{\text{-3}}$) and very recently has been reported to be associated with a globular cluster in the M81 spiral galaxy system at a distance of 3.6\,Mpc~\citep{Kirsten+2021b}. 
If this association is correct, FRB~20200120E would be the nearest extragalactic FRB.  At a distance that is $\sim$40 times closer than the next closest FRB at 149~Mpc \citep{Marcote+2020}, FRB~20200120E can be used to explore burst luminosities $\gtrsim 10^3$ times lower than other FRBs.

Multi-frequency follow-up studies are critical for characterizing  FRB sources and emission models. One such approach is to carry out simultaneous observations of FRBs across wide bandwidths \citep{Scholz+2017, Majid+2020, Pearlman+2020, Scholz+2020} that are more robust against temporal evolution of scintillation and scattering, as well as intrinsic short term variability in the emission spectrum of individual bursts. Another approach is to probe the shorter timescales of the bursts to place limits on the instantaneous size of the emitting regions, similar to the studies of the Crab giant pulses carried out with few nanosecond time resolution \citep{Hankins+2003}. Observations carried out at higher radio frequencies are particularly robust against temporal scattering due to multi-path propagation that can limit the overall effective time resolution.

In this Letter, we present results from a simultaneous observation of FRB~20200120E at 2.25 and 8.36\,GHz with the NASA Deep Space Network~(DSN) 70\,m telescope, DSS-63, located in Madrid, Spain. The observation and data analysis procedures are described in Section~\ref{Section:Observation}. In Section~\ref{Section:Results}, we provide measurements of the brightest burst detected, including the DM, width, flux density, fluence, and burst morphology. In Section~\ref{Section:Discussion}, we discuss our measurements of the burst spectra, implications of the narrow components, burst energetics, and the impact of intrinsic and extrinsic effects on the burst properties.


\section{Observations and Data Analysis}
\label{Section:Observation}

We observed FRB~20200120E continuously for 4.1\,hr, starting at 2021~March~2 21:35:20~UTC (MJD~59275.89953), using DSS-63, the NASA DSN 70\,m radio telescope located at the Madrid Deep Space Communication Complex~(MDSCC) in Robledo, Spain. 
This observation was performed using the pointing position RA~(J2000)\,$=$\,09:57:56.688, Dec~(J2000)\,$=$\,+68:49:31.800. 
We carried out this observation as  part of a monitoring program of repeating FRBs at high frequencies with the DSN's large 70\,m radio telescopes, making use of the available gap times in the DSN's schedule for radio astronomical observation of FRBs. DSS-63 is equipped with cryogenically-cooled, dual circular polarization receivers, which are capable of recording data simultaneously at $S$-band and $X$-band. The center frequencies of the recorded $S$-band and $X$-band data were 2.25 and 8.36\,GHz, respectively. The $S$-band system has an effective bandwidth of 115\,MHz, after taking into account the front-end filter roll-off and masking bad channels contaminated by radio frequency interference~(RFI). The $X$-band receivers provide 450\,MHz of usable bandwidth. Data from both polarization channels were simultaneously received and recorded at each frequency band with two different recorders at the site's Signal Processing Center. One of the recorders is the \text{wide-band} pulsar machine, capable of recording four independent input bands simultaneously, each with a bandwidth of 500 MHz with a frequency resolution of 0.5 MHz and time resolution of 2.2 ms. In addition, we used a set of baseband recorders to record both S-band and X-band data in two polarization modes with time resolution of 62.5 nanoseconds, albeit at the expense of sacrificing some portion of the available bandwidth. The analysis described in this Letter used data from the baseband recorders, and as such we will describe the configuration of these recorders in some detail below. The three baseband recorders were configured with a variable number of 16\,MHz sub-bands, with two-bit per sample recording. One recorder was used to record both polarization channels at S-band, with 7 sub-bands for each polarization channel, spanning \text{2197--2309} MHz contiguously.  The second and third baseband recorders were used to record the two polarization channels at X-band, using 16 sub-bands for each polarization channel, spanning \text{8200--8456} MHz.

The initial data processing procedures were similar to those described in previous studies of the Crab giant pulses with the DSN \citep[e.g.,][]{Majid+2011}. A nominal DM value of 87.82\,pc\,cm$^{\text{-3}}$, as initially reported by \citet{Bhardwaj+2021}, was used to coherently dedisperse the data in each \text{sub-band}. The power in each \text{sub-band's} time series was summed over all \text{sub-bands} to form one time series per polarization mode. A list of burst candidates with detection signal-to-noise~(S/N) ratios above 7.0 were generated using a matched filtering algorithm, in which each dedispersed time-series was convolved with boxcar functions ranging in widths from 62.5 nanoseconds up to 16.2 ms, in multiplicative factors of 2. For the bright burst described in this Letter, we subsequently extracted a few seconds of data around the nominal burst time and coherently dedispersed that data over a range of DM values to obtain an optimal DM value of 87.77(2)\,pc\,cm$^{\text{-3}}$ while maximizing the signal-to-noise~(S/N) of the burst. Using the updated DM value, we then proceeded to correct the data for the bandpass slope across the frequency band. We also subtracted the moving average from each data value using 0.5\,s around each time sample to remove the low frequency temporal variability. The data samples were then normalized by the off-burst standard deviation, before summing the two polarization bands to form total power spectra.

The data in each observing band were flux calibrated using nominal $T_{\text{sys}}$ values	published by the 
DSN\footnote{See \href{https://deepspace.jpl.nasa.gov/files/DSN_Radio_Astronomy_Users_Guide.pdf} {https://deepspace.jpl.nasa.gov/files/DSN\_Radio\_Astronomy\_Users\_Guide.pdf.}}, which we previously verified via ON--OFF observations of bright 3C calibrators, such as 3C123, 3C48, and Cas~A. The $T_{\text{sys}}$ values were corrected for elevation effects, which are minimal for elevations greater than 20 degrees. We conservatively estimate the error on flux densities in each band to be 15\%, or less.


\section{Results}
\label{Section:Results}

We detected several bursts from FRB~20200120E during this 4.1~hr observation. Among these, one burst stood out as the brightest with very detailed structure down to the native time resolution of the observation. In this Letter, we describe the properties of this burst, hereafter referred to as B1, and defer the presentation of the remaining bursts and their attributes to a later publication. 

\begin{figure*}[b]
	\centering
	\includegraphics[trim=0cm 0cm 0cm 0cm, clip=false, scale=0.7, angle=0]{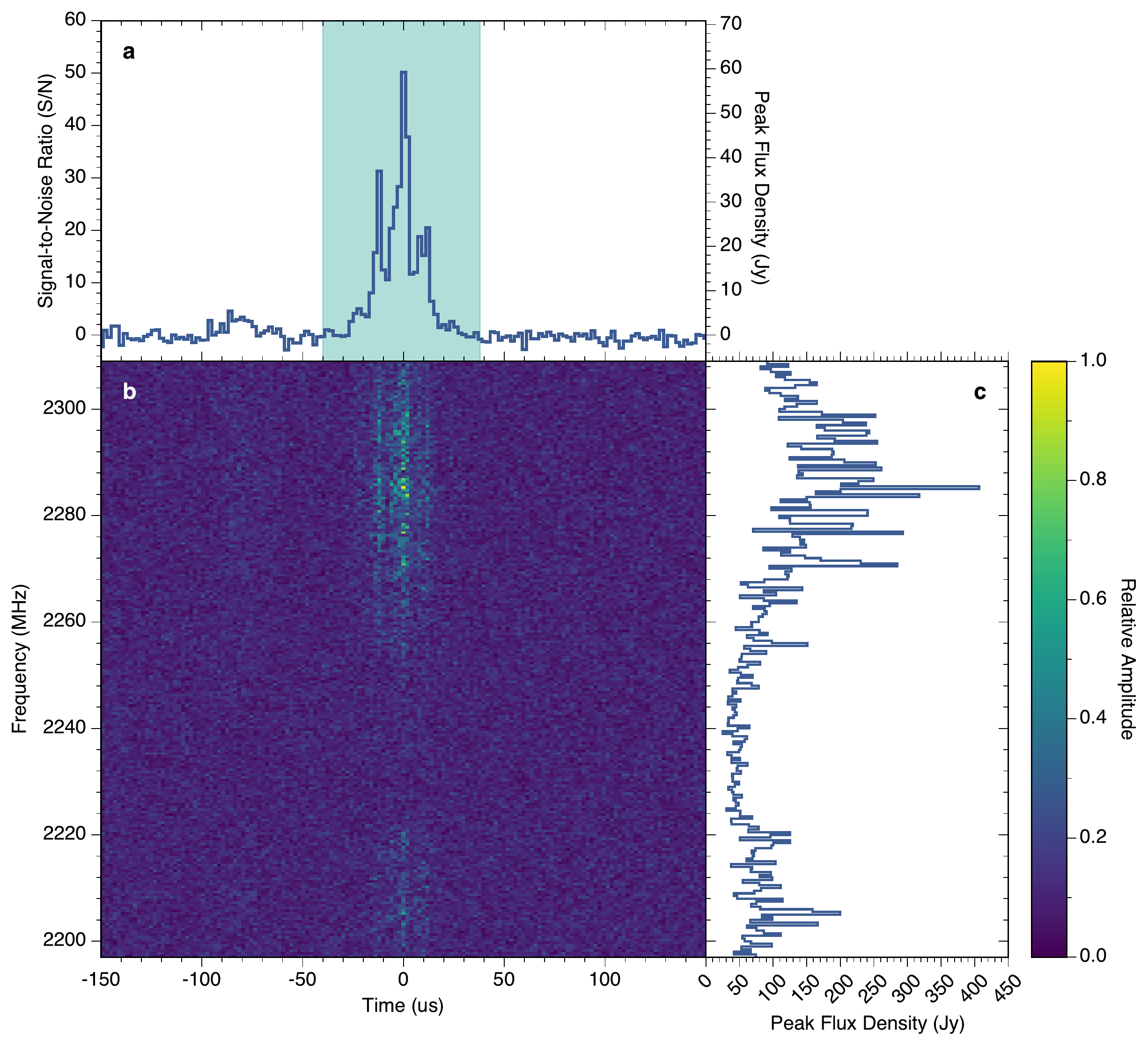}
	\caption{Dynamic spectrum of B1, the brightest $S$-band burst detected during our observation. The data are plotted with a time and frequency resolution of 2\,$\mu$s and 0.5\,MHz, respectively. The S/N maximizing DM (87.77\,pc\,cm$^{\text{--3}}$) was used for dedispersion. Panel (a) shows the frequency-averaged, dedispersed burst profile. The teal shaded region shows the interval that was used for computing the on-pulse spectrum in panel (c). The dedispersed dynamic spectrum is shown in panel (b), and the color bar indicates the relative intensity of the features.}
	\label{Figure:Figure1}
\end{figure*}

In Figure~\ref{Figure:Figure1}, we show the flux-calibrated dedispersed $S$-band dynamic spectrum of the burst B1 using a DM value of 87.77\,pc\,cm$^{\text{--3}}$.
In this case, we coherently dedispersed the baseband data to form filterbanks comprised of channelized power  spectral  densities  with  temporal  and  spectral  resolutions  of  2 $\mu$s  and 0.5 MHz,  respectively.  
The frequency-averaged burst profile is shown in the upper panel, while the time-averaged on-pulse spectrum is shown in the right panel.

\begin{figure*}[b]
	\centering
	\includegraphics[trim=0cm 0cm 0cm 0cm, clip=false, scale=0.9, angle=0]{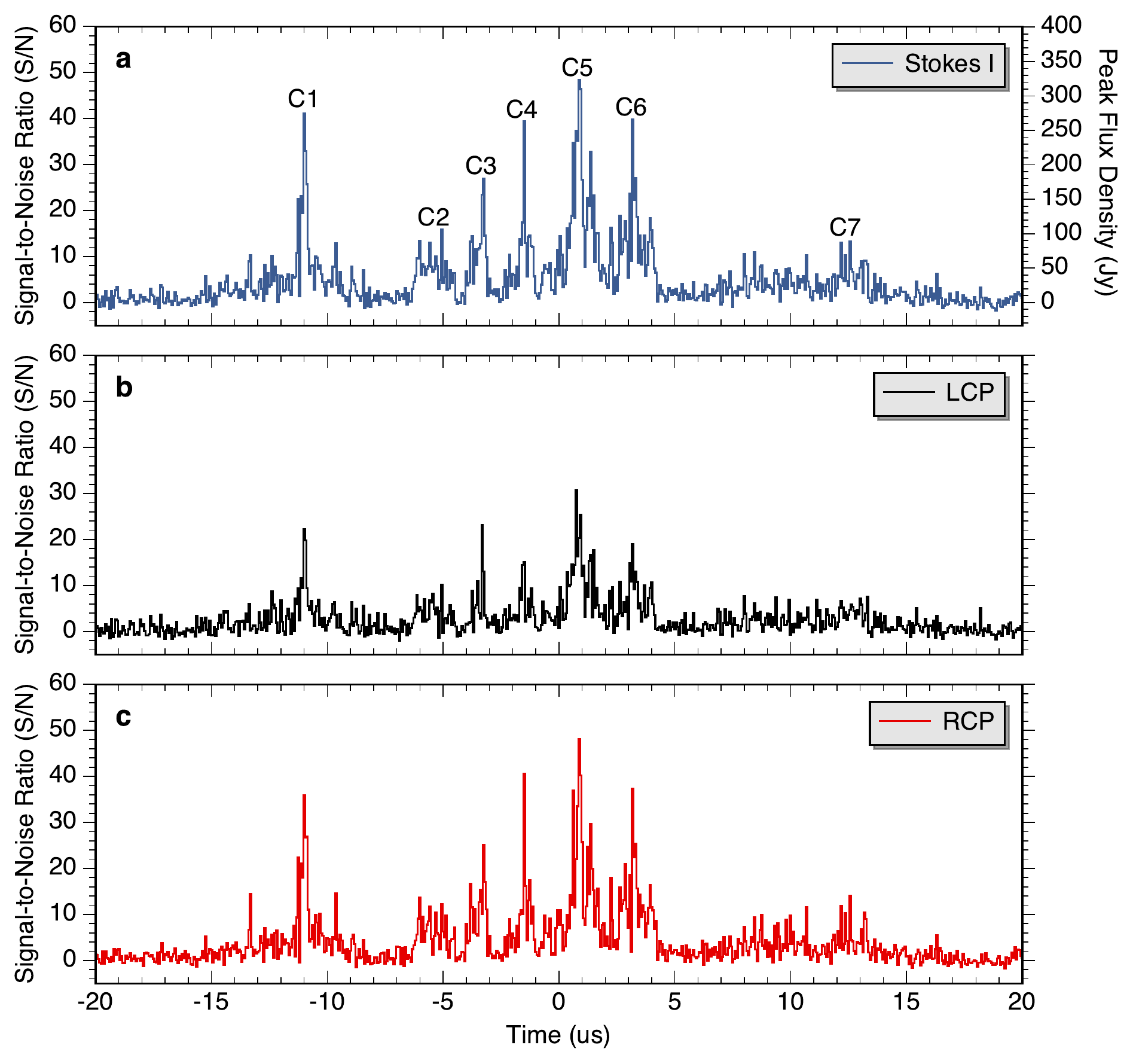}
	\caption{Coherently dedispersed $S$-band burst profiles shown at 62.5 ns time resolution, derived from baseband data with subbands of 16~MHz. In panel (a), we show the Stokes~I burst profile, derived from combining the~LCP and RCP~baseband data. The LCP and RCP burst profiles are shown in panels~(b) and~(c), respectively. The data have been dedispersed using the S/N maximizing DM of 87.77\,pc\,cm$^{\text{--3}}$. This plot shows that the burst is comprised of multiple bright, narrow peaks, with widths of $\lesssim$\,100\,ns. The labels~(C1--C7) displayed in panel~(a) correspond to seven distinct pulse components comprising the burst.}
	\label{Figure:Figure2}
\end{figure*}

In Figure~\ref{Figure:Figure2}, we show the coherently dedispersed 
(87.77\,pc\,cm$^{\text{--3}}$) burst time series with 62.5~ns time 
resolution.  
The top panel shows the burst in flux-calibrated total intensity, while the lower panels show the right and left circular polarizations given in units of S/N. A detailed polarization analysis of B1 and additional bursts from this observation will be presented in a forthcoming publication. 

The burst is made up of several remarkably 
narrow, isolated components, numbering at least seven non-overlapping 
components, labeled as C1-C7 in Figure~\ref{Figure:Figure2}. 
Component C1 is relatively narrow and appears to be preceded by a short period of low-level emission.
This component is followed by a series of similarly narrow components with typical separation of 2-3 $\mu$s.
Component C4 is not resolved temporally with a width that is at least as short as the sampling time of 62.5 ns.
 The last component, C7, seems to be broader with a width of a few $\mu$s.
 We also note the presence of a broad precursor emission in Figure~\ref{Figure:Figure1}, arriving $\sim80 \mu$s prior to the arrival time of B1.
In Table~\ref{Table:Table1}, we 
list the peak time, peak S/N, DM value that maximized the peak S/N, 
burst width, burst peak flux density, and fluence. We also show in this 
table the flux density and fluence of the narrowest component of the burst, C4.

Although B1 was detected with high S/N at $S$-band, there was no detectable signal during the same time at $X$-band. Since no bursts were detected at $X$-band, we place a 7-$\sigma$ upper limit of 1.1\,Jy on the flux density of the emission at 8.36\,GHz during this observation, assuming a similar burst width of 33\,$\mu$s. At the narrowest component width (62.5 ns), corresponding to a single time sample in this observation, the 7-$\sigma$ upper limit on the flux density at $X$-band is 25\,Jy.

If we further assume that the flux density of FRBs scales as a power-law (i.e., $S(\nu)$\,$\propto$\,$\nu^{\alpha}$, where $S(\nu)$ denotes the flux density at an observing frequency $\nu$ and $\alpha$ is the spectral index), which is typical of most pulsar radio spectra, then we can place an upper limit of $\alpha$\,$<$\,--1.3 on the spectral index of the emission process for the wider burst. On the other hand, using the measured flux density of the narrowest component at $S$-band and assuming its width at $X$-band is similarly narrow, we can  place an upper limit of $\alpha$\,$<$\,--1.8 on the spectral index of the emission process. However, we note that previous multi-wavelength observations of FRBs show patchy emission over large, instantaneous frequency bandwidths \citep{Majid+2020, Pearlman+2020} and do not appear to support a power-law model~(e.g.,~\citealt{Spitler+2016, Scholz+2016, Law+2017}).

The burst spectrum (Figure~\ref{Figure:Figure1}) shows frequency structure 
on the order of 10~MHz across the band.  Fitting Gaussians to the two 
bumps in the spectrum, we find full-width half-max values of 
${\rm FWHM}_1 = 39.1 \pm 0.7$~MHz at $\nu_{c,1} = 2286.6 \pm 0.3$~MHz and 
${\rm FWHM}_2 = 24 \pm 2$~MHz at $\nu_{c,2} = 2207.5 \pm 0.8$~MHz.  This 
structure could be intrinsic or a result of diffractive interstellar 
scintillation.  The NE2001 electron density model of the Milky Way 
\citep{Cordes+2002} predicts 
that our Galaxy will impart a pulse broadening time of 
$\tau_{\rm sc, 1} = 200$~ns for an extragalactic source along the line of 
sight to this FRB ($\ell = 142\fdg2$, $b=+41\fdg2$).  Scaling to 2.3~GHz 
as $\nu^{-4}$, this gives $\tau_{\rm sc} = 8$~ns, well below the time 
resolution of our data.  The scintillation bandwidth corresponding to this 
scattering time is 
\begin{equation}
\Delta \nu_{\text{DISS}} \sim \frac{1}{2\pi\tau_{\rm sc}}.
\label{Equation:DISS}
\end{equation}
which gives $\Delta \nu_{\text{DISS}} \approx 20$~MHz at 2.3~GHz.  The 
frequency structure we see in our burst spectrum is entirely consistent 
with modulation from scintillation, but measurements at other frequencies 
will be needed to confirm this.  
We note that the spectrum of B1 is not well modeled by a power law (i.e., $S(\nu)$\,$\propto$\,$\nu^{\alpha}$, where $S(\nu)$ denotes the flux density at an observing frequency $\nu$ and $\alpha$ is the spectral index). This is also confirmed in previous studies of FRB spectra, where generally the burst spectra are not well modeled by a power law dependence (e.g., ~\citealt{Scholz+2016, Law+2017, Majid+2020}).
If the structure is indeed caused by 
scintillation, then this could also explain the lack of a detection at 
8.36~GHz where the scintillation bandwidth would be 
$\Delta \nu_{\text{DISS}} \approx 3.5$~GHz and we could have $100\%$ 
modulations in our band.  However, FRBs are typically not broadband, so 
the lack of a detection at $X$-band could also be explained by the intrinsic 
structure of the burst.
Observations at other frequencies, particularly instantaneous broadband observations, will confirm whether the frequency structure we see in B1 at S-band along with a simultaneous lack of emission at X-band is indeed the result of scintillation or intrinsic  structure of the burst.  Similar narrow-band emission behavior has also been in seen in FRB 121102 and FRB 180916.J0158+65 , where detection of bright emission at lower frequencies have not been accompanied with emission at higher frequencies during simultaneous observations (\citealt{Majid+2020, Scholz+2020, Pearlman+2020}).


\begin{deluxetable}{lc}
	\tablenum{1}
	\tabletypesize{\small}
	\tablecolumns{2}
	\tablewidth{0pt}
	\tablecaption{\textsc{Radio Burst B1 from FRB~20200120E Detected with DSS-63}}
	\startdata
	Peak Time~(MJD) & 59275.99804894308$^{\mathrm{a,g}}$ \\
	DM & 87.77\,$\pm$\,0.02$^{\mathrm{b}}$ \\
	Peak S/N & 50.1$^{\mathrm{c,g}}$ \\
	Burst Width & 33\,$\pm$\,1\,$\mu$s$^{\mathrm{d,g}}$ \\
	Burst Peak Flux Density & 59\,$\pm$\,12 Jy$^{\mathrm{e,g}}$ \\
	Burst Fluence & 0.76\,$\pm$\,0.15\,Jy\,ms$^{\mathrm{e,f,g}}$ \\
	Flux Density of Narrowest Component~(C4) & 270\,$\pm$\,54\,Jy$^{\mathrm{e,h}}$ \\
	Fluence of Narrowest Component~(C4) & 0.017\,$\pm$\,0.003\,Jy\,ms$^{\mathrm{e,f,h}}$
	\enddata
	\tablecomments{\\
		$^{\mathrm{a}}$ Barycentric time of arrival~(ToA) using the position RA~(J2000)\,$=$\,09:57:56.7, Dec~(J2000)\,$=$\,68:49:32.0~\citep{Bhardwaj+2021}. \\
		$^{\mathrm{b}}$ DM value that maximized the peak S/N of the burst. \\
		$^{\mathrm{c}}$ Peak signal-to-noise ratio~(S/N) after dedispersing with the peak~S/N-maximizing DM of 87.77\,pc\,cm$^{\text{--3}}$. \\
		$^{\mathrm{d}}$ Burst width at 10 percent maximum.\\
		$^{\mathrm{e}}$ Uncertainties are dominated by the 20\% fractional error on the system temperature, $T_{\text{sys}}$. \\
		$^{\mathrm{f}}$ Fluence determined using full width at 10\% of peak SNR. This choice ensures that all of the burst energy is included. \\
		$^{\mathrm{g}}$ Value is derived from data sampled at a time resolution of 2\,$\mu$s. \\
		$^{\mathrm{h}}$ Value is derived from data sampled at a time resolution of 62.5\,ns.}
	\label{Table:Table1}
\end{deluxetable}


\section{Discussion}
\label{Section:Discussion}

High time resolution studies of	FRBs are an important observational tool for discerning proposed emission mechanisms for FRBs, improving our ability to measure fundamental properties of FRBs such as burst energetics, shortest timescales of the emission process that
could constrain the spatial sizes of the emission region, and burst energy densities. Furthermore, such high time resolution studies make it possible to carry out detailed studies of scatter broadening due to the turbulent plasma between the source and the observer, including	studies	of ISM in the Milky Way, the Milky Way halo, host galaxy halo, ISM in the host galaxy, and finally any local scattering material near the progenitors of FRBs. Finally, high time resolution studies of FRBs will also help in	identifying known analogs to FRB emission characteristics, such	as the giant-pulse phenomenon seen in some pulsars, Type III solar radio bursts,	and short S-shaped decametric Jovian bursts \citep[see, e.g.,][]{Hankins+2015, Tan+2019, Shaposhnikov+2011}.

\subsection{Burst Microstructure}
\begin{figure*}[b]
	\centering
	\includegraphics[trim=0cm 0cm 0cm 0cm, clip=false, scale=0.7, angle=0]{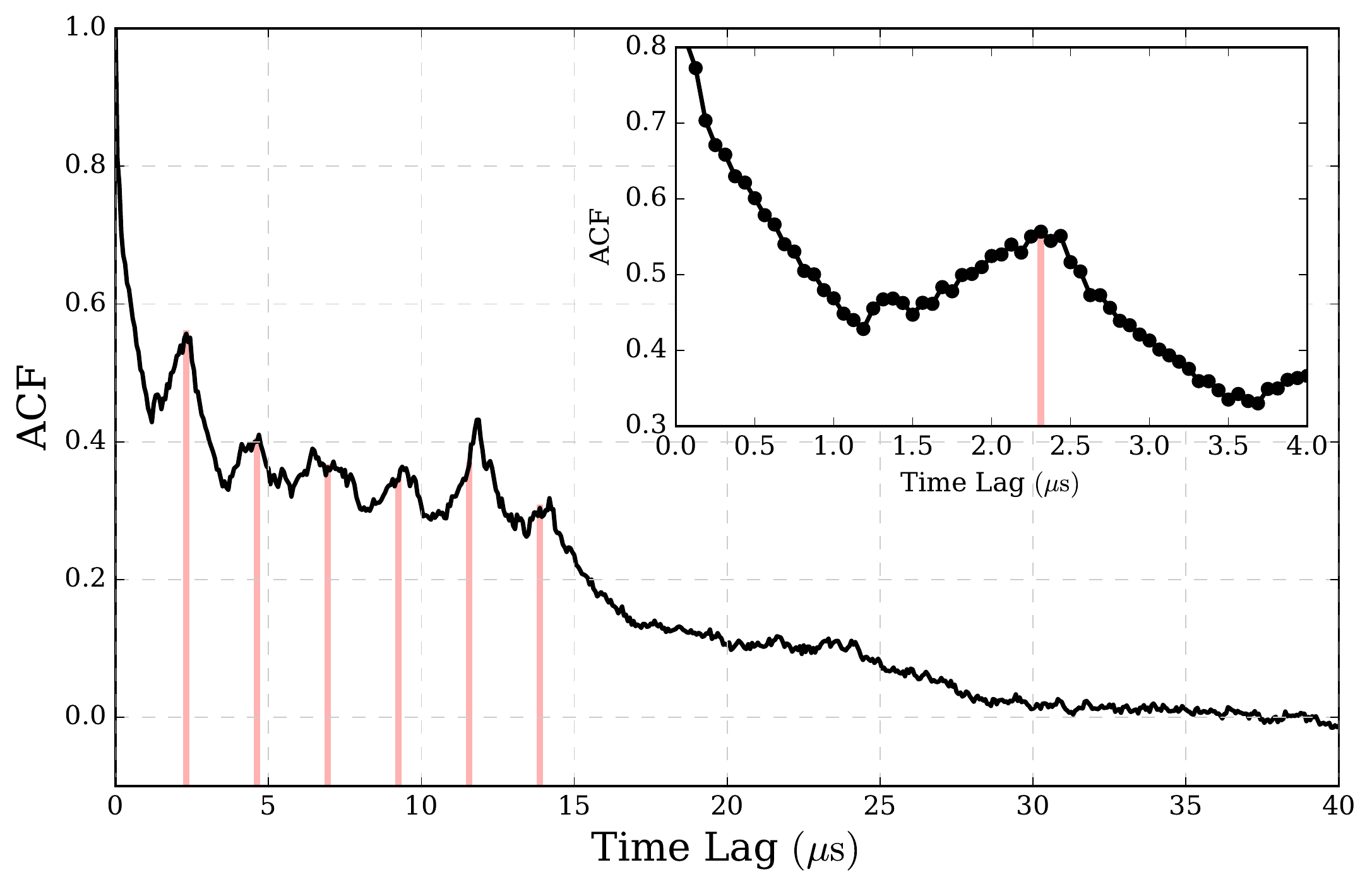}
	\caption{Autocorrelation function of the total intensity burst 
	         out to time lags of 40~$\mu$s.  At time lags 
	         $\lesssim 15~{\rm \mu s}$, there are six peaks 
	         with roughly equal spacing.  The red lines show integer 
	         multiples of 2.3~$\mu$s, which is the time lag of the 
	         first peak.  The last peak is at a lag of 
	         $\approx 14~{\rm \mu s}$, which corresponds to the 
	         time between component 1 and component 6 in 
	         Figure~\ref{Figure:Figure2}.  The inset shows the ACF 
	         at small lags, and the red line indicates the maximum of 
	         the first peak.
	         }
	\label{Figure:Figure3}
\end{figure*}

As shown in Figure~\ref{Figure:Figure2}, B1 is comprised of several 
narrow ($\lesssim1~{\rm \mu s}$) components contained within a larger 
burst envelope ($\sim 33~{\rm \mu s}$).  In Figure~\ref{Figure:Figure3}, 
we show the autocorrelation function (ACF) of the total intensity burst B1.  
The ACF shows correlations out to time lags of $\approx 30~{\rm \mu s}$ 
and six peaks at time lags $\lesssim 15~\mu$s.  The peaks in the ACF have 
widths of 1--2~$\mu$s and are roughly evenly spaced with separations of 
$\approx 2.3~\mu$s.  This is consistent with what we see in B1, where the 
seven components shown in Figure~\ref{Figure:Figure2} have separations of 
1.9--2.7~$\mu$s (or integer multiples thereof).  
While the component separations are not strictly periodic in B1, there is 
clearly a repetition timescale of $\approx2~\mu$s in the burst structure.  

The structure seen in B1 is qualitatively similar to what was seen in the 
scattering-limited observations of FRB~20180916B \citep{Nimmo+2021}, but on much longer time scales, 
and is strikingly similar to what is seen in some giant pulses from the 
Crab pulsar \citep{Hankins+2003, Hankins+2007}. 
While the exact emission mechanism may not be the same between FRBs and 
Crab giant pulses, bursts from both sources are generically expected to be 
comprised of many individual shots of emission \citep{Cordes+2016}.  
The small scattering towards FRB~20200120E and the high time resolution 
of these observations have enabled the clear detection of burst 
microstructure.



The roughly periodic occurrence of burst components within B1 is also 
reminiscent of the microstructure seen in single pulses of some radio 
pulsars.  Pulsar microstructure describes the rapid intensity fluctuations 
within individual single pulses.  ACF analyses like the one performed here 
have revealed that the microstructure of some pulsars has recurring 
periodicities \citep[e.g.,][]{Cordes+1990,Lange+1998}.  While not seen 
in every radio pulsar, microstructure periodicities have been measured 
in several slow spinning pulsars \citep{Lange+1998, Mitra+2015}, the young Vela 
pulsar \citep{kramer-2002}, and even millisecond pulsars \citep{De+2016}.
Interestingly, there is a correspondence between the microstructure 
periodicity ($P_\mu$) and the pulsar spin period ($P_{\rm spin}$) given 
by

\begin{equation}
P_\mu \sim 10^{-3} P_{\rm spin}
\end{equation}

albeit with significant scatter \citep{De+2016,Mitra+2015}.  

While it is not known if the origin of pulsar microstructure is geometric (beam sweeping past) 
or temporal modulation,
the phenomenological 
similarity between B1 and periodic microstructure in pulsars is tantalizing 
and may suggest a magnetospheric origin for FRB~20200120E.
If the $P_\mu \approx 2-3~{\rm \mu s}$ structure seen in B1 follows the 
$P_\mu - P_{\rm spin}$ relation for pulsar microstructure, it would imply 
a spin period of several milliseconds, which could be achieved by a fast-spinning 
young pulsar or a recycled millisecond pulsar.  The latter would be consistent with
the localization of this FRB to a globular cluster \citep{Kirsten+2021b}, which tend to 
be fertile grounds for the formation of millisecond pulsars.

\subsection{Luminosity and Brightness Temperature}
\begin{figure*}[b]
	\centering
	\includegraphics[trim=0cm 0cm 0cm 0cm, clip=false, scale=0.7, angle=0]{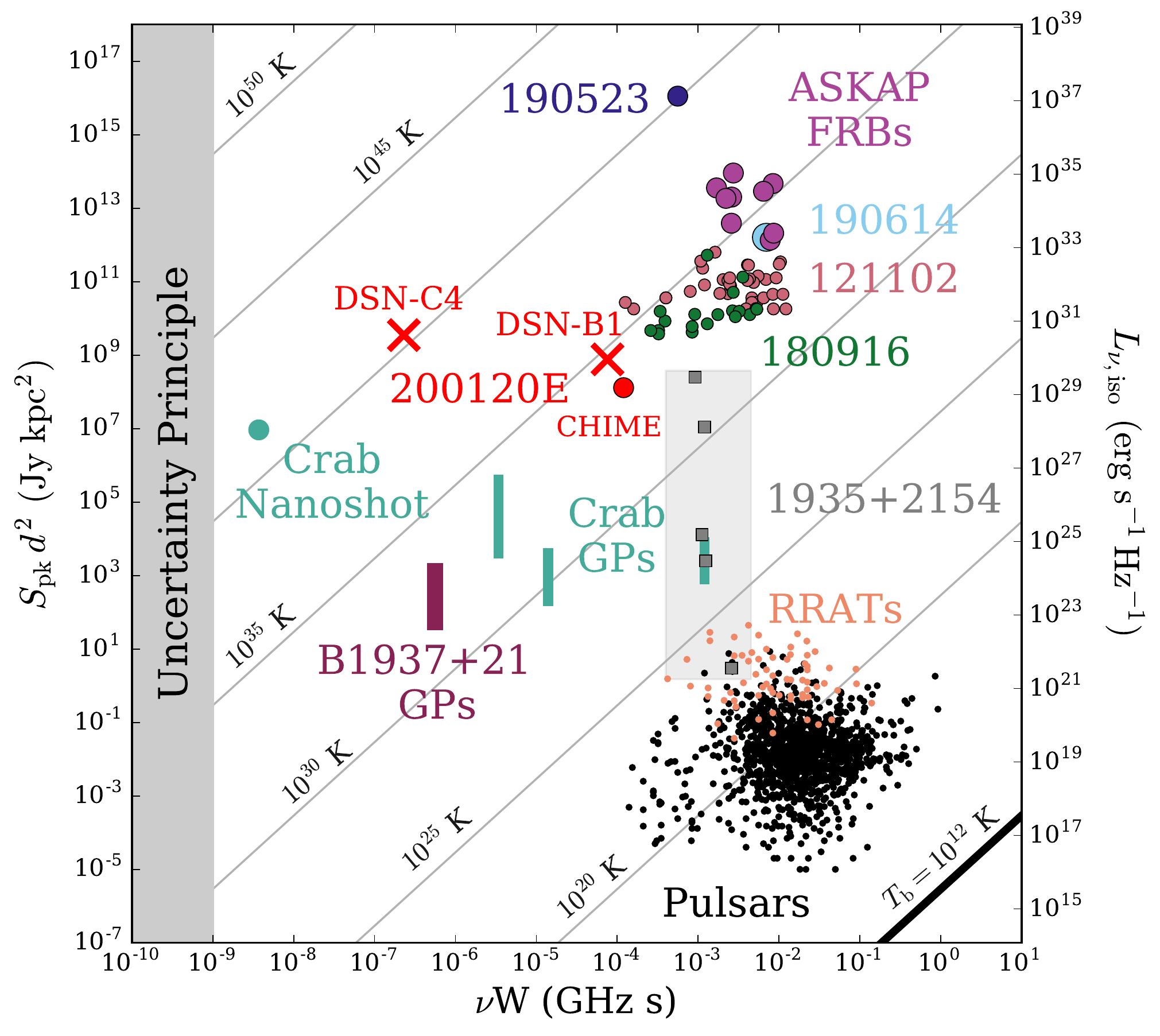}
	\caption{Time-luminosity phase-space plot for radio bursts showing 
	         the product of observing frequency and pulse width ($\nu W$ in 
	         GHz~s) against the pseudoluminosity ($S_{\rm pk} d^2$, 
	         left) and equivalent isotropic spectral luminosity 
	         ($L_{\nu, \rm iso} = 4\pi S_{\rm pk} d^2$, right).  
	         The brightness temperature 
	         $T_{\rm b} = (1/2k_{\rm b}) \times (Sd^2) / (\nu W)^2$
	         is indicated by black lines with the nominal limit for 
	         coherent emission ($T_{\rm b} \sim 10^{12}~{\rm K}$) 
	         shown as a thick black line.  The shaded grey region 
	         at $\nu \rm W \lesssim 10^{-9}$~GHz~s is excluded due 
	         to the uncertainty principle.
	         Pulsars \citep{Manchester+05} and RRATs with known 
	         distances are shown in black and orange dots, respectively. 
	         Giant pulses from the Crab pulsar \citep{Lundgren+1995, Karuppusamy+2010, Majid+2011} and B1937+21 \citep{McKee+2019} 
	         are shown as green and purple lines, respectively, that 
	         indicate the range of pulse flux densities observed.  
	         The 0.4~ns Crab ``nanoshot'' \citep{Hankins+2007} is 
	         shown as a green dot.  Bursts from the magnetar 
	         SGR~1935+2154 are shown as grey squares in a shaded grey 
	         box that indicates the extent of observed emission 
	         \citep{Bochenek+20, chime-sgr2020, Kirsten+2021a, Zhang+2020}.  
	         FRBs with host galaxies with measured redshifts are also shown 
	         and labeled.  For the repeaters FRB~121102 \citep{Spitler+2016,Scholz+2016,Gajjar+2018} and 
	         FRB~180916 \citep{Marthi+2020,Marcote+2020}, multiple 
	         bursts are shown.  For FRB~200120E, we show the initial 
	         CHIME detection as a red circle \citep{Bhardwaj+2021}, 
	         and our results as red crosses.  DSN-B1 is the 
	         full 33~$\mu$s burst and DSN-C4 is the narrowest 
	         ($\sim 100$~ns) component of the burst.
	         }
	\label{Figure:Figure4}
\end{figure*}

Figure~\ref{Figure:Figure4} shows how the full burst B1 and its narrowest 
component C4 compared to other short duration radio pulses in the time-luminosity 
phase space.  If we assume that the burst source is at the same distance as M81 
($d = 3.6$~Mpc), then B1 is on the faint end of the FRB luminosity distribution 
with a pseudo-luminosity of 
$S_{\rm pk} d^2 \approx 8\times 10^8~{\rm Jy~kpc}^2$ and an isotropic equivalent 
spectral luminosity of 
$L_{\rm \nu, iso} = 4\pi S_{\rm pk} d^2 \approx 9 \times 10^{29}~{\rm erg~s}^{-1}~{\rm Hz}^{-1}$, 
which is only a factor of $\approx 3$ larger than the SGR~1935+2154 burst 
detected by STARE2 \citep{Bochenek+20}.  Continued monitoring of this FRB 
with large telescopes will characterize the faint end of the FRB luminosity 
distribution to levels below the STARE2 burst.

The narrowest component, C4, has a pseudo-luminosity of  
$S_{\rm pk} d^2 \approx 3.5\times 10^9~{\rm Jy~kpc}^2$ and an 
isotropic equivalent spectral luminosity of 
$L_{\rm \nu, iso} \approx 4 \times 10^{30}~{\rm erg~s}^{-1}~{\rm Hz}^{-1}$.
This is only $\approx\! 380$ times larger than the Crab nanoshot 
($S=2.3$~MJy, $W=0.4$~ns, $d=2$~kpc) seen by \citet{Hankins+2007} 
and occupies the sparsely populated short duration region of the 
time-luminosity phase space (Figure~\ref{Figure:Figure4}).  The brightness temperature, given by 
\begin{equation}
T_{\rm b} = \frac{S d^2}{2 k_{\rm B} \left(\nu W\right)^2}
\label{Equation:bright_temp}
\end{equation}
is $T_{\rm b} \approx 2 \times 10^{40}$~K for component C4, which   
is one of the highest measured $\rm{T_{b}}$ values for any FRB and only a 
factor of $\approx 10$ smaller than the Crab nanoshot \citep{Hankins+2007}.  
Of course, since this narrow component is still unresolved 
in time, the true brightness temperature may be much larger.

\subsection{Size of the Emission Region}
\label{sec:timescale}

\cite{Farah+2018} reported the discovery of FRB~170827, which shows complex temporal structure with components as short as $\sim$30 $\mu$s, indicating that the emission region may be as small as only 10 km. Similarly, \cite{Cho+2020} showed the presence of microstructure in their study of FRB~181112, resolving emission components as short at 15 $\mu$s, implying emission size of a few km. \cite{Nimmo+2021} reported on high time resolution observations of FRB~20180916B, where they observed several bursts with a range of timescales from a few ms to as short as a few $\mu$s wide. Their observations, carried out at 1.7 GHz, were limited in resolving narrower structures by multi-path scattering. They attribute the observation of 3--4 $\mu$s structure in one of the detected bursts to an emission region on the order of 1 km.

We are reporting a significantly shorter variability time scale for FRB~20200120E than for any previously published FRB. 
Very recently \cite{Nimmo+2021b} also reported similar timescale emission for FRB~20200120E.
While the envelope of emission spans $\sim \mathrm{30 \mu s}$, the burst is clearly comprised of distinct and non-overlapping narrow components.
Using a light travel time argument, the short duration of the individual components imply small source sizes $\sim \mathrm{c\,\delta t}$, where $\mathrm{\delta t}$ is the width of individual components. For an emission timescale of $\sim$ 100 ns, corresponding to the width of the narrowest component, the size of the emission region is constrained to be $\mathrm{r \lesssim 30\,m}$. 

However, relativistic effects can significantly alter the relation between size of the emission region and the variability time scale. For instance,\ \cite{beniamini+kumar20} consider the case of a relativistic outflow moving with a Lorentz factor $\gamma$ producing a radio burst at distance $R$ from the origin of the outflow, and define a characteristic time scale, $t_0$, 

\begin{equation}
t_0(R) \equiv \frac{R}{2 c \gamma^2} \sim 70\, {\rm ns}\ \left( \frac{R}{10^{9}\ {\rm cm}}  \right) \ \left( \frac{\gamma}{500} \right)^{-2} 
\label{eq:tchar}
\end{equation}

\noindent
where the scaling has been chosen to give a timescale comparable to the shortest time scale observed in the burst reported here.
Large $\gamma$ or relatively small $R$ are required to explain variability time scales
with $t_0 \lesssim 100 {\rm ns}$.
As discussed by \cite{beniamini+kumar20}, the variability time scale may be even shorter than the characteristic time scale given in Eq, \ref{eq:tchar} due to a number of possible factors: clumps in the emission region, a narrow range of emission radii, radial evolution of the spectrum, or anisotropic emission.  Given our observation that the burst components appear to have comparable rise and fall times, clumpiness or anisotropic emission would be favored over narrow emission radii or radial spectral evolution.
In any case, as \cite{beniamini+kumar20} point out, the efficiency of conversion of energy from the relativistic outflow must decrease when a physical mechanism is invoked to decrease the time scale of variability.

\subsection{Energy Density}

Using the light travel time estimate, and combining the size of the emission region with the measured fluence (Table 1), we estimate the energy spectral density of the component to be $\mathrm{0.15 \times 10^{13} J\,m^{-3}\,Hz^{-1}}$ in the rest frame of the emitting source. Under the conservative assumption that the emission is confined only to our observing band, we estimate the total energy density of the component to be $\mathrm{1.6 \times 10^{20} J\,m^{-3}}$.  If we attribute the entire energy density to the equivalent energy density in a magnetic field $\mathrm{u_{B} = B^2/2\mu}$, where B is the magnetic field and $\mu$ is the magnetic permeability, we obtain a magnetic field $\mathrm{B \sim 2 \times 10^{11} G}$.  
 We note that if relativistic effects such as beaming are taken into account (see the discussion in \ref{sec:timescale} above)  the estimated magnetic field required to serve as the emission reservoir can be significantly lower than the value estimated above.


\subsection{Propagation Effects}
The extremely high time resolution of our data allows us to 
set a pulse broadening limit of $\tau_{\rm sc} \lesssim 100$~ns 
to the total 2.3~GHz scattering along the line of sight to 
FRB~20200120E.  Scaling to 1~GHz as $\nu^{-4}$, this corresponds to a limit 
of $\tau_{\rm sc,1~GHz} \lesssim 3~\mu{\rm s}$.  A small scattering 
time is to be expected since the line of sight ($\ell = 142\fdg2$, 
$b=+41\fdg2$) only passes through a small part of the Milky Way 
and the FRB likely resides in the outer edges of M81.  

The NE2001 electron density model \citep{Cordes+2002} predicts a 
Milky Way contribution to the 2.3~GHz pulse broadening time of 
$\tau_{\rm sc} = 8$~ns at 2.3~GHz along the line of sight to the FRB.  
Although this is well below our time resolution limit, the corresponding
scintillation bandwidth is $\Delta \nu_{\rm DISS} \approx 20$~MHz, 
which is entirely consistent with the frequency structure we see in 
B1.  If there is any additional scattering caused by the host 
galaxy or the Milky Way halo (which is not modelled in NE2001), 
it would be limited to $\tau_{\rm sc} \lesssim 100$~ns, which would 
correspond to a scintillation bandwidth of 
$\Delta \nu_{\rm DISS} \gtrsim 2$~MHz at 2.3~GHz.   
We see no obvious signs of correlations on frequency 
separations less than $\sim10$~MHz, which suggests that any 
additional scattering is smaller than the contribution from the 
Milky Way disk.  This is consistent with the apparent location of 
the FRB outside M81 \citep{Bhardwaj+2021} and the small amount of 
scattering expected from the Milky Way halo \citep{Ocker+2021}.

Observations at other frequencies will confirm whether the 
frequency structure we see in B1 is indeed the result 
of scintillation.  Since the line of sight appears to be 
relatively uncontaminated by both the host and our own Galaxy,
this FRB could be a very nice source for probing the properties 
of the ionized gas in the Galactic halo.  Furthermore, the 
very small scattering seen here ($\tau_{\rm sc} \lesssim 100$~ns) 
means that even at 100~MHz the scattering would be 
$\tau_{\rm sc, 0.1} \lesssim 30$~ms.  We therefore strongly 
encourage low frequency observations of FRB~20200120E for probing
the Galactic halo.

\subsection{Future Outlook}

With the detection of unresolved burst structure at the	native time resolution (62.5\,ns) during the observations reported here, the question remains whether the individual components in bursts from FRB~20200120E are even narrower in width at nanosecond timescales. Such studies are best carried out at higher radio frequencies, where the effects of multi-path scatter broadening are mitigated. Since the temporal width of the bursts is a fundamental observational property of the emission processes and has important implications for determining the burst energetics, we encourage future observations of FRBs with even higher temporal resolutions at higher radio frequencies. On the other hand, as mentioned in the previous section, FRB~20200120E also offers an opportunity to study the properties of the ionized medium in the halo of the Milky Way Galaxy. Low frequency observations, where the effects of multi-path scattering are amplified, would also be quite useful for this purpose.


\section*{Acknowledgments}

We thank Paz Beniamini and Sterl Phinney for informative discussions during the preparation of this paper.

A.B.P is a McGill Space Institute (MSI) Fellow and a Fonds de Recherche du Quebec -- Nature et Technologies (FRQNT) postdoctoral fellow. R.S.W. is supported by an appointment to the NASA Postdoctoral Program at the Jet Propulsion Laboratory, administered by Universities Space Research Association under contract with NASA. M.B. is supported by an FRQNT Doctoral Research Award. C.L. was supported by the U.S. Department of Defense~(DoD) through the National Defense Science and Engineering Graduate Fellowship~(NDSEG) Program. E.P. acknowledges funding from an NWO Veni Fellowship.

We thank the Jet Propulsion Laboratory's and California Institute of Technology's President's and Director's Research and Development Fund for supporting this work. We also thank Dr. Charles Lawrence and Dr. Stephen Lichten for providing programmatic support. In addition, we are grateful to the DSN scheduling team (Hernan Diaz, George Martinez, Carleen Ward), the Madrid Deep Space Communication Complex~(MDSCC) staff for scheduling and carrying out these observations, and Steven Olson for his help in flux calibration of the data.

A portion of this research was performed at the Jet Propulsion Laboratory, California Institute of Technology and the Caltech campus, under a Research and Technology Development Grant through a contract with the National Aeronautics and Space Administration. U.S. government sponsorship is acknowledged.

FRB research at UBC is supported by an NSERC Discovery Grant and by the Canadian Institute for Advanced Research.




\newpage


\bibliographystyle{yahapj}
\bibliography{references}


\end{document}